\documentclass[a4paper,12pt]{article}

\usepackage[centertags]{amsmath}
\usepackage{amsfonts}
\usepackage{amssymb}
\usepackage{amsthm}
\usepackage{newlfont}
\usepackage[english]{babel}
\usepackage{enumerate}

\newtheorem{theorem}{Theorem}

\newtheorem{lemma}{Lemma}

\theoremstyle{remark}

\theoremstyle{definition}

\theoremstyle{definition}

     \newcommand {\beq}  {\begin{equation}}
      \newcommand {\eeq}  {\end{equation}}

\author{ Lakshtanov E.L.\thanks{Department of
Mathematics, Aveiro University, Aveiro 3810, Portugal.  This work
was supported by {\it Centre for Research on Optimization and
Control} (CEOC) from the ''{\it Funda\c{c}\~{a}o para a
Ci\^{e}ncia e a Tecnologia}'' (FCT), cofinanced by the European
Community Fund FEDER/POCTI, and by the FCT research project
PTDC/MAT/72840/2006.} \thanks{e-mail: lakshtanov@rambler.ru}}

\title{Spectral properties of the Dirichlet-to-Neumann operator for exterior Helmholtz problem and its applications to scattering theory}

\begin{document}
\date{}
\maketitle

\begin{abstract}
We prove that  the Dirichlet-to-Neumann operator (DtN) has no
spectrum in the lower half of the complex plane. We find several
application of this fact in scattering by obstacles with impedance
boundary conditions. In particular, we find an upper bound for the
gradient of the scattering amplitude and for the total cross
section. We justify numerical approximations by providing bounds
on difference between theoretical and approximated solutions
without using any a priory unknown constants.
\end{abstract}

\section{Introduction}
In this article we discuss some spectral properties of the so
called Dirichlet-to-Neumann map which allows to determine many
properties of the scattering amplitude for scattering by obstacles
with impedance boundary conditions. We remind the reader that,
up to now, we did not have any concrete information on the
scattering properties for obstacles of arbitrary shape in case of
intermediate values of the frequency. Besides,  in numerical
schemes (like Galerkin's scheme, for example) all inequalities
controlling the difference between theoretical and constructed
solutions include some, a priori, unknown constant,  which depends
on the surface. Our results on the spectrum of DtN allow to exclude
this dependence.

The article has the following structure. First, we prove absence
of the spectrum in the lower halfspace for the operator
DtN. Then,  theorem 2 states an upper bound for the difference between
theoretical and approximated solution. Theorem 3 lists upper
bounds for total cross section, gradient of the scattering
amplitude, field on the boundary and its normal derivative. And
finally, theorem 4 is a note on the wave analogue of the Newton's
minimal resistance problem, namely we present a lower bound for
the transport cross section.

Consider a bounded body $\Omega \subset \mathbb R^3$ with smooth
boundary $\partial \Omega$ and $k > 0$. The scattered field is
given by the Helmholtz equation and a radiation condition
\begin{equation}\label{helm}
\Delta u(r)+k^2 u(r)=0, \quad r \in \Omega'=\mathbb R^3 \backslash
\Omega,
\end{equation}
\begin{equation}\label{Somm}
\int_{|r|=R} \left |\frac{\partial u(r)}{\partial |r|}-iku(r)
\right |^2 dS = o(1), \quad R \rightarrow \infty,
\end{equation}
If we  fix the quite smooth boundary condition on $\partial
\Omega$,
\begin{equation}\label{bcOmega}
u(r)=u_0(r), \quad u_0 \in W_2^{1/2}(\partial \Omega),
\end{equation}
then there exists a unique solution which satisfies all these conditions (eg
\cite{AlRamm}). Every function  $u(r)$ which satisfies the
mentioned conditions has asymptotic
\begin{equation}\label{scamp}
u(r)=\frac{e^{ik|r|}}{|r|} u_{\infty}(\theta)+o \left
(\frac{1}{|r|} \right ), \quad r \rightarrow \infty, \quad
\theta=r/|r| \in S^2,
\end{equation}
where the function $u_\infty(\theta)=u_\infty(\theta,k,u_0)$ is
called the {\it scattering amplitude} and the quantity
$$
\sigma_{u_0}=\|u_\infty\|^2_{L_2(S^2)}=\int_{S^2}
|u_\infty(\theta)|^2 d\mu (\theta)
$$
is called the total cross section. $\mu$ is a square element of
the unit sphere.

The operator $F$ which associates a boundary condition $u_0 \in
 C(\partial \Omega)$ to the scattering amplitude $u_\infty$ is called the
 Far field operator. Its boundedness easily follows from the existence of the Dirichlet Green function
 \cite{AlRammPhysA},\cite{AlRammGr}, therefore,  it can be continued to a bounded operator $F :
L_2(\partial \Omega, dS) \rightarrow L_2(S^2,d\mu)$, where $dS$ is
a standard square measure on $\partial \Omega$.

The operator $DtN : L_2(\partial \Omega) \rightarrow L_2(\partial
\Omega)$ associates a function $u_0$  to the normal derivative of
the corresponding field $u(r)$.
$$
DtN(u_0) = \frac{du}{dn}(r), \quad r \in \partial \Omega.
$$
Operator $DtN$ with domain $\{u_0 \in W^2_{1/2}(\partial \Omega) :
DtN u_0 \in L_2(\partial \Omega)\}$  is unbounded,
pseudodifferential operator of order 1 with compact resolvent
\cite[Ch.7]{tay},
 \cite{uhlmann},\cite{vainberg},\cite[Th 3.11]{coltonKress}.

\begin{theorem}\label{first}
The operator $DtN$ has no spectrum in the lower
half of $\mathbb C$.
\end{theorem}
Note that in the case where $\partial \Omega$ is a sphere this
fact was known earlier (eg \cite{ned}).

{\bf Proof}. Let us prove that for every function $u \in
C^2(\mathbb R^3 \backslash \Omega) \cap C^1(\overline{\mathbb R^3
\backslash \Omega})$ which satisfies (\ref{helm}),(\ref{Somm}) we
have
 \begin{equation}\label{spectr}
\left  \| \frac{\partial u}{\partial n}+(a+ib)u \right
\|_{L_2(\partial \Omega)} \geq b \| u \|_{L_2(\partial \Omega)},
\end{equation}
where $a,b \in \mathbb R, \quad b>0$.
$$
\left \| \frac{\partial u}{\partial n}+(a+ib)u \right \|^2 = \left
\| \frac{\partial u}{\partial n}+a u \right \|^2+2b \Im \left (
\int_{\partial \Omega} \frac{\partial u}{\partial n} \overline{u}
dS \right )+b^2 \|u\|^2.
$$
The proof is finished by the well know fact (which follows  from the Second
Green's identity)
\begin{equation}\label{sigma} \Im \left ( \int_{\partial \Omega}
\frac{\partial u}{\partial n} \overline{u} dS \right )=k \|
u_\infty \|^2_{L_2(S^2)} \geq 0,
\end{equation}
Note now that the the inverse to $\left[ DtN+(a+ib) \right ]$ is
defined on a dense set in $L_2(\partial \Omega)$, since the boundary
problem (\ref{helm}),(\ref{Somm}) and
$$
(DtN+a+ib)u=f,
$$
are uniquely solvable for   $f \in W^{1/2}_2(\partial \Omega)$
(eg. \cite{AlRamm}). According to (\ref{spectr}), it is bounded on
this set and therefore can be continually extended to a bounded
operator acting on $L_2(\partial \Omega)$. The theorem is proved.

\subsection{Justification of arbitrary numerical schemes with uniform constant}
Let the field $u^\gamma$ satisfy conditions
(\ref{helm}),(\ref{Somm}) and impedance boundary
conditions of the form
\begin{equation}\label{dirA}
\left ( \frac{\partial }{
\partial n}+\gamma(r) \right )u^\gamma  \equiv f(r) , \quad
\quad r \in
\partial \Omega, \quad f \in C(\partial \Omega).
\end{equation}
where the function $f(r)$ is supposed to be known, the impedance
function $\gamma(r) \in C(\partial \Omega)$ has  positive
imaginary part
 $\Im(\gamma(r)) \geq \gamma_0 >0$ and $\gamma_0$ is a
constant.
The
existence and uniqueness of the solution of
(\ref{helm}),(\ref{Somm}),(\ref{dirA}) is proven, for example, in
\cite{vantychonovsamarsky,AlRamm}.

Suppose that we have found a function $u^{\gamma,1}$  that satisfies
(\ref{helm}),(\ref{Somm}) and almost satisfies (\ref{dirA}) (it is
not important how it was found, either by applying numerical
schemes or using analytical  approximations in case of small or
large values of $k$):
\begin{equation}\label{adir}
\left ( \frac{\partial }{
\partial n}+\gamma(r) \right )u^{\gamma,1}  \equiv f(r)
+\alpha(r), \quad \alpha(r) \in L_2(\partial \Omega,dS).
\end{equation}
In what follows, $\| \cdot \| = \| \cdot \|_{L_2(\partial
\Omega)}$ and  $\Gamma=\| \gamma\|_{C(\partial \Omega)}$.
\begin{theorem}\label{third}
1. We have an upper bound for the difference of fields:
\begin{equation}
 \| u^\gamma-u^{\gamma,1}\| \leq \frac{1}{\gamma_0 }\| \alpha \|
\end{equation}
2. There is an upper bound for the difference of normal derivatives:
\begin{equation}
\left  \| \frac{\partial}{\partial
n}u^\gamma-\frac{\partial}{\partial n}u^{\gamma,1} \right \| \leq
 \left (\frac{\Gamma}{\gamma_0 }+1 \right )\| \alpha \|
\end{equation}
3. And finally, there is an upper bound for the difference between total cross
sections of theoretical and constructed waves:
\begin{equation}
\left  \| u^\gamma_\infty-u^{\gamma,1}_\infty \right
\|^2_{L_2(S^2)} \leq  \frac{1}{k\gamma_0}\left
(\frac{\Gamma}{\gamma_0 }+1 \right )\| \alpha \|^2
\end{equation}
\end{theorem}
The proof is given in part \ref{proof3}.

\subsection {Scattering of a plane wave by obstacle with impedance
boundary conditions}  Now we consider  scattering of the incident
field $e^{ik(r \cdot \theta_0)}$  formed by a plane wave with
incident angle $\theta_0 \in S^2$, by an obstacle $\Omega$. Let
the field $u^\gamma$ satisfy conditions (\ref{helm}),(\ref{Somm})
and impedance boundary conditions of the form
\begin{equation}\label{dir}
\mathcal B_\gamma (u^\gamma)|_{\partial \Omega} \equiv - \mathcal
B_\gamma (e^{ik(r \cdot \theta_0)})|_{\partial \Omega} , \quad
\quad r=(x,y,z) \in
\partial \Omega, \quad
\end{equation}
where  $\gamma(r) \in C(\partial \Omega)$ is a positive function
such that $\Im(\gamma(r)) \geq \gamma_0 >0$, $\gamma_0$ is a
constant, and $\mathcal B_\gamma =(\partial /
\partial n)+k\gamma(r)$.
The operator $B_{i\gamma}$ appears as a stationary analogue of the
$\frac{\partial}{
\partial n} -  \gamma(r) \frac{\partial}{\partial t}$ for the time-dependent wave equation.

\begin{theorem}\label{second}
Let $u^{\gamma}$ satisfy (\ref{helm}),(\ref{Somm}),(\ref{dir}).
Denote by $S$ the area of  $\partial \Omega$. The
following inequalities hold:

1. We have an upper bound for the total cross section:
\begin{equation}\label{f_3}
\sigma_\gamma =\| u^\gamma_{\infty} \|^2_{L_2(S^2)} \leq S
\frac{(1+\Gamma)^2(\gamma_0+\Gamma)}{\gamma_0^2}
\end{equation}
2. We have an upper bound for the gradient of the scattering amplitude:
\begin{equation}\label{f_4}
| \nabla_\theta u^\gamma_{\infty}(\theta) |  \leq \frac {\sqrt{S}
k}{4\pi} \frac{1+\Gamma}{\gamma_0} (k(\gamma_0+\Gamma)+k+1), \quad
\theta \in S^2 \subset \mathbb R^3.
\end{equation}

3. And finally there are bounds for the field and normal
derivative of the field on the surface $\partial \Omega$.
 \begin{equation}\label{f1} \|u^\gamma\| \leq \sqrt{S}
\frac{1+\Gamma}{\gamma_0},
\end{equation}
4.
\begin{equation}\label{f2}
\left \| \frac{\partial u^\gamma}{\partial n} \right \| \leq k
\sqrt{S}\frac{(1+\Gamma)(\gamma_0+\Gamma)}{\gamma_0}
\end{equation}
\end{theorem}
In case of constant value of $\gamma(r)$ the first statement was
proven in \cite{lc}. The $2^{nd}$ inequality is a consequence of
the well known representation:
\begin{equation}\label{amplRep}
u_\infty(\theta)=\frac{1}{4\pi} \int_{\partial \Omega} \left (
\frac{\partial u}{\partial n}+i k (n \cdot \theta) u \right ) e^{-ik(\theta \cdot r)} dS(r).
\end{equation}
We should note, that we used everywhere  $|(n \cdot \theta)| \leq
1$, so  these inequalities could be improved.



\section{Wave analogue of the Newton's problem of
body minimal resistance}  In 1685 Newton published
\cite{Principia} the solution of his problem of minimal
resistance. The body flies through a rarefied medium where
particles do not mutually interact and have elastic collisions
with the body's surface. Newton considered convex bodies of
revolution embedded in a certain cylinder and having the same
geometrical cross section $\sigma_{cl}$. He obtained  an exact
positive solution in this case. Recently, a body with zero
resistance  was constructed \cite{inv}. It is  interesting  to
study the problem of minimization of the resistance of a body in
case of wave scattering.

In the wave model of  scattering by an obstacle, the observable
corresponding to classical resistance, is the transport cross section
(eg \cite{sph})
$$
R_\gamma(k,\theta_0,\Omega) = \int_{S^2} (1-(\theta \cdot
\theta_0)) |f_\gamma(\theta)|^2 d\theta.
$$
Note that by definition the resistance is normalized by the total
cross section $R_\gamma /  \sigma_\gamma \in [0,2]$ and
clearly, in classical scattering, the border values of the segment
$[0,2]$  can be attained. Of course,
since the distribution of the scattered wave $f(\theta)$ is an
analytical function,  it can not equal a $\delta$-function and so
$R_\gamma$ cannot be equal  to zero. But, due to the
quasiclassical effect, for large obstacles (or wave numbers) the
infimum of the normalized $R_\gamma \backslash \sigma_\gamma$
could be $0$. To see this effect, one can fix $k$ and observe a
sequence of prolate spheroids ($a=b=n$, $c=1 \backslash n, n
\rightarrow \infty)$. By the results of \cite{sph} we have for
every convex body
$$
\lim_{k \rightarrow \infty} \frac{R_\infty}{\sigma_\infty}  = \frac{R_{cl}}{\sigma_{cl}}
$$
and this ratio could become arbitrary small in our sequence of spheroids.
\begin{theorem}\label{newton}
The following inequality holds
\begin{equation}\label{trcs}
 R_\gamma > \frac{1}{2\pi} \left (
\frac{\sigma_\gamma \gamma_0}{k S} \right )^2
\frac{1}{(1+\Gamma)^2(1+\Gamma+\gamma_0)^2} , \quad \Im \gamma>0.
\end{equation}
\end{theorem}
Therefore, we can conclude that $R_\gamma$ has a positive infimum in the
class of obstacles with fixed total cross section $\sigma_\gamma$
and uniformly bounded area $S$.

\section{Discussion of the results}
1. Inequality (\ref{f_3}) (Theorem 2, part 1.) solves the question
whether for certain $\gamma$ and $k>0$ there exists a sequence of smooth
obstacles with uniformly bounded area such that $\sigma_{\gamma}$
 tends to infinity.

Note that this fact is quite nontrivial, since plane waves transfer
infinite energy and every part of it interacts with the obstacle,
even if it is quite far from the obstacle.

2. There exist many numerical methods of obstacle reconstruction
from scattering data. But (\ref{f_3}) gives us  the possibility to
estimate the area of the obstacle immediately, since we measured
the scattering amplitude for any body angle. See
\cite{coltonPiana} for another approach.


3. Inequality (\ref{f_4}) (Theorem 3 part 2) gives us the
possibility to extrapolate values of the scattering amplitude in
case it is only known on the some net.

4. Theorem 2 evidently tells us exactly when we have to stop our
numerical scheme.

\section{Proofs of results}\label{proof3}

In what follows, $\widehat{\Gamma}=\|\Im (\gamma)\|_{C(\partial
\Omega)}$.
\begin{lemma}\label{general} For every field $u \in
C^2(\mathbb R^3 \backslash \Omega) \cap C^1(\overline{\mathbb R^3
\backslash \Omega})$ which satisfies
(\ref{helm}),(\ref{Somm}),(\ref{bcOmega})
\begin{equation}\label{lm1}
\left  \| \frac{\partial u}{\partial n}+k\gamma(r) u \right \|
\geq \gamma_0 k \| u \|.
\end{equation}
\end{lemma}
{\bf Proof.}
$$
\left  \| \frac{\partial u}{\partial n}+k\gamma(r) u \right \| =
\left  \| \frac{\partial u}{\partial n}+k
Re(\gamma)u+ik\widehat{\Gamma} u
+ik(\Im(\gamma(r))-\widehat{\Gamma}) u \right \| \geq
$$
$$
\left \| \frac{\partial u}{\partial n}+k
Re(\gamma(r))+ik\widehat{\Gamma} u\right \|-
\|k(\Im(\gamma(r))-\widehat{\Gamma}) u \| \geq
$$
$$
k \widehat{\Gamma} \|u\| - k \| \Im(\gamma(r))-\gamma_0 \|_C \|u\|
\geq k \gamma_0 \|u\|.
$$
Here we used that $\| \Im(\gamma(r))-\gamma_0 \|_C \leq
\widehat{\Gamma} - \gamma_0.$ The Lemma is proved.
\\
\\

Let us prove theorem \ref{third}. Using lemma \ref{general}, we get
$$
 \gamma_0 \| u^\gamma-u^{\gamma,1}\| \leq \left \| \left ( \frac{\partial }{
\partial n}+\gamma(r) \right )
(u^\gamma-u^{\gamma,1}) \right \| = \|\alpha\|
$$
2. Using $\left ( {\partial } \slash {
\partial n}+\gamma(r) \right )(u^\gamma-u^{\gamma,R})=\alpha(r)$, we
get
$$
\left \| \frac{\partial}{\partial
n}u^\gamma-\frac{\partial}{\partial n}u^{\gamma,1} \right \| \leq
\Gamma \|u^\gamma-u^{\gamma,1} \| + \|\alpha \| \leq  \left
(\frac{\Gamma}{\gamma_0 }+1 \right )\| \alpha \|
$$
3. Using (\ref{sigma})
$$
\| u^\gamma_\infty-u^{\gamma,1}\infty\|^2_{L_2(S^2)} \leq
\frac{1}{k} \|u^\gamma-u^{\gamma,1} \| \cdot \left \|
\frac{\partial}{\partial n}u^\gamma-\frac{\partial}{\partial
n}u^{\gamma,1} \right \| \leq
$$
$$
\frac{1}{k} \frac{1}{\gamma_0 }\| \alpha \|  \left
(\frac{\Gamma}{\gamma_0 }+1 \right )\| \alpha \| =
\frac{1}{k\gamma_0}\left (\frac{\Gamma}{\gamma_0 }+1 \right )\|
\alpha \|^2
$$
Theorem \ref{third} is proven.

Now, let us prove theorem \ref{second}. Note  that from (\ref{dir}), it
follows that
\begin{eqnarray*}
\left \|\frac{\partial u^{\gamma}}{\partial n}+k\gamma(r)
u^{\gamma} \right \| &=& \left \|\frac{\partial e^{ik(r \cdot
\theta_0)}}{\partial n}+k\gamma(r)
e^{ik(r \cdot \theta_0)} \right \| \leq\\
& \leq & \left \| \frac{\partial e^{ik(r \cdot
\theta_0)}}{\partial n} \right \| +k\Gamma \|e^{ik(r
\cdot \theta_0)}\| \leq \sqrt{S}k(1+\Gamma)\\
\end{eqnarray*}
Recall that $S=Area(\partial \Omega)$. Hence, using (\ref{lm1}),
we obtain
\begin{equation}\label{ff2}
 \gamma_0 \|u^{\gamma}\| \leq \sqrt{S} (1+\Gamma)
\end{equation}
Also from (\ref{dir}), we have
$$
-\frac{\partial u^{\gamma}}{\partial n}=k\gamma(r) u^{\gamma}(r) +
\frac{\partial e^{ik(r \cdot \theta_0)}}{\partial n}+k\gamma(r)
e^{ik(r \cdot \theta_0)},
$$
therefore
\begin{equation}\label{lastlast}
\left \|\frac{\partial u^{\gamma}}{\partial n} \right \| \leq
k\Gamma \|u^{\gamma}|+ \left \| \frac{\partial e^{ik(r \cdot
\theta_0)}}{\partial n} \right \| + k\Gamma\|e^{ik(r \cdot
\theta_0)}\| \leq
\end{equation}
$$
k\Gamma \|u^{\gamma}\|+\sqrt{S}k(1+\Gamma) \leq
{2k\sqrt{S}(1+\gamma)}
$$
Now from (\ref{ff2}) and (\ref{lastlast}), we have
\begin{equation}\label{alm1}
\sigma_{\gamma}\leq \frac{1}{k} \|u^{\gamma}\| \|\frac{\partial
u^{\gamma}}{\partial n}\| \leq
 \frac{1}{k} \left (\frac{\sqrt{S} (1+\Gamma)}{\gamma_0}
\right )\left (2{k\sqrt{S}(1+\Gamma)} \right )=
\frac{{2S}(1+\Gamma)^2}{\gamma_0}.
\end{equation}
This ends the proof of the theorem \ref{second}.

Now we prove theorem \ref{newton}. From (\ref{amplRep}), we obtain
the upper bound for the scattering amplitude for every angle:
\begin{equation}\label{amplb}
|f(\theta)|\leq \frac{1}{4\pi} \left ( \left \| \frac{\partial
u_\gamma}{\partial n} \right \| + k \| u_\gamma \| \right )
\sqrt{S} =\frac{kS}{4\pi \gamma_0} (1+\Gamma)(\gamma_0+\Gamma+1)
=: M, \quad \theta \in S^2.
\end{equation}
$$
R_\gamma=\int_{0}^{\pi}\int_0^{2\pi} (1-\cos \widetilde{\theta})
|f(\widetilde{\theta},\varphi)|^2 d(-\cos \widetilde{\theta}) d
\varphi \geq
$$
$$
\int_{\widetilde{\theta} : 1 -\cos \widetilde{\theta} > \delta
}^{}\int_0^{2\pi} (1-\cos \widetilde{\theta})
|f(\widetilde{\theta},\varphi)|^2 d(-\cos \widetilde{\theta}) d
\varphi,
$$
where $1 \geq \delta \geq 0$ is an arbitrary number. Using
(\ref{amplb}) we obtain that the last expression is greater than
$$
\delta \left ( \sigma_\gamma - \int_{\widetilde{\theta} : 1 -\cos
\widetilde{\theta} < \delta }^{}\int_0^{2\pi}
|f(\widetilde{\theta},\varphi)|^2 d(-\cos \widetilde{\theta}) d
\varphi\right ) \geq \delta (\sigma_\gamma - 2\pi \delta M^2).
$$
Choosing $\delta:=\frac{\sigma_\gamma}{4\pi M^2}$ we obtain
$$
R_\gamma \geq \frac{\sigma_\gamma^2}{8 \pi M^2} =\frac{1}{2\pi}
\left ( \frac{\sigma_\gamma \gamma_0}{k S} \right )^2
\frac{1}{(1+\Gamma)^2(1+\Gamma+\gamma_0)^2} .
$$
Theorem \ref{newton} is proven.


\begin{thebibliography}{102}
\bibitem{vantychonovsamarsky} A.N.Tychonov, A.A.Samarsky, ``Equations
of Mathematical Physics Pergamon", Oxford, (1963).

\bibitem{coltonKress} David L. Colton, Rainer Kress, Inverse Acoustic and Electromagnetic Scattering
Theory, Series: Applied Mathematical Sciences  , Vol. 93, 1998.


\bibitem{coltonPiana} David Colton, Michele Piana, Inequalities
for inverse scattering problems in absorbing media, Inverse
Problems, 17 (2001), 597-605.

\bibitem{AlRamm} A.G.Ramm, ``Scattering by Obstacles" (Dordrecht:
Reidel), (1986)




\bibitem{majda} A.~Majda, ``High frequency Asymptotics for the Scattering matrix and
the inverse problem of Acoustical scattering", \emph{Comm. pure
and applied math.} vol. XXIX, 261--291, (1976)


\bibitem{sph} A.I.Aleksenko, W. de Roeck, E.L.Lakshtanov, ``Resistance of the Sphere to a Flow of Quantum Particles'', J.Phys.
A.Math.Gen, (39), pp. 4251-4255, 2005.


\bibitem{lc}   A.Aleksenko, P. Cruz, E. Lakshtanov, ''High-frequency limit of the transport cross section in scattering by an obstacle with impedance boundary conditions'',
 2008 J. Phys. A: Math. Theor. 41 255203

\bibitem{wl} W. de Roeck, E.L.Lakshtanov, ``Total cross section exceeds transport cross section for
quantum scattering from hard bodies at low and high wave numbers'', J.Math.Phys, 48, 2007.



\bibitem{majdaTay} A.~Majda, M.E.Taylor, ``The asymptotic behavior of the diffractive peak in classical scattering'', ~
\emph{Comm. pure and applied math.} vol. XXX, 639--669, (1977)


\bibitem{AlRammPhysA} S. Gutman, A.G. Ramm, Numerical implementation of the MRC method for obstacle scattering
problems, J. Phys. A: Math. Gen. 35 (2002) 8065-8074

\bibitem{AlRammGr} S. Gutman, A.G. Ramm, Modified Rayleigh Conjecture Method and Its
Applications, arXiv:math/0601298v1 [math.NA]

\bibitem{CAP} Ramm. A.G., Calculation of the scattering amplitude for the wave scattering
from small bodies of an arbitrary shape, Radiofisika, 12, (1969),
1185-1197. 43,7131.

\bibitem{Principia}
I. Newton,\, {\it Philosophiae naturalis principia mathematica},\,
1686.

\bibitem{inv}
Alena Aleksenko, Alexander Plakhov, {\it Bodies of zero resistance
and bodies invisible in one direction}, Nonlin 22(6), pp.
1247-1258, 2009.


\bibitem{uhlmann} G. Uhlmann, Inverse boundary value problems and applications,
Asterisque, 207 (1992), pp. 153-211.

\bibitem{tay} M.E. Taylor, Partial Differential Equations II: Qualitative
Studies of Linear Equations. Springer-Verlag, New-York. 1996.

\bibitem{ned} J.-C. N\'{e}d\'{e}lec, Acoustic and Electromagnetic Equations, Integral
Representations for Harmonic Problems, Springer-Verlag, 2001.

\bibitem{vainberg} B. R. Vainberg and V. Grushin, Uniformly nonelliptic problems,
Math. USSR-Sbornik, 2 N1 (1967), pp. 111-133.

\end{thebibliography}
\end{document}